%% file: paper.tex
\documentclass[conference]{IEEEtran}
\IEEEoverridecommandlockouts
\usepackage{cite}
\usepackage{amsmath,amssymb,amsfonts}
\usepackage{algorithmic}
\usepackage{graphicx}
\usepackage[caption=false,font=footnotesize]{subfig}
\usepackage{textcomp}
\usepackage{xcolor}
\usepackage{circuitikz}
\usepackage{dblfloatfix}
\def\BibTeX{{\rm B\kern-.05em{\sc i\kern-.025em b}\kern-.08em
    T\kern-.1667em\lower.7ex\hbox{E}\kern-.125emX}}
\begin{document}


\title{Active Filter Design for Buffering Datacenter-Scale Power Fluctuations from Training AI\\
\thanks{This work was supported by the Bits \& Watts Initiative at Stanford University.}
}

\author{
\IEEEauthorblockN{Dillon Jensen}
    \IEEEauthorblockA{
    \textit{Stanford University}\\
    Stanford, CA, USA \\
    dillonj0@stanford.edu}
\and
\IEEEauthorblockN{Grant Wilkins}
    \IEEEauthorblockA{
    \textit{Stanford University}\\
    Stanford, CA, USA \\
    gfw@stanford.edu}
\and
\IEEEauthorblockN{Obi Nnorom Jr.}
    \IEEEauthorblockA{
    \textit{Stanford University}\\
    Stanford, CA, USA \\
    obdk@stanford.edu}
\and
\IEEEauthorblockN{Hugo Budd}
    \IEEEauthorblockA{
    \textit{Stanford University}\\
    Stanford, CA, USA \\
    hugorbudd@gmail.com}

\and

\IEEEauthorblockN{Phil Levis}
    \IEEEauthorblockA{
    \textit{Stanford University}\\
    Stanford, CA, USA \\
    pal@cs.stanford.edu}
\and
\IEEEauthorblockN{Ram Rajagopal}
    \IEEEauthorblockA{
    \textit{Stanford University}\\
    Stanford, CA, USA \\
    ramr@stanford.edu}
\and
\IEEEauthorblockN{Juan Rivas}
    \IEEEauthorblockA{
    \textit{Stanford University}\\
    Stanford, CA, USA \\
    jmrivas@stanford.edu}
}





\maketitle

\begin{abstract}
Training large AI models requires thousands of processors working in parallel. This synchronized load poses a challenge for traditional power delivery architectures, because IT power ramps much faster than grid hardware can compensate.

This paper presents a hardware solution to this problem, introducing a novel power delivery architecture which automatically detects changes in load power and compensates using on-rack energy storage, giving time for the grid to respond to the varying load. The architecture is validated experimentally, powering a training load while limiting the grid-side power ramp rate such that it stays within pre-specified range. The prototype presented is rated to deliver up to 10 kW of buffered power in a 400 VDC system, with a bill of materials cost of \$3,500 USD. The system design is described in detail. A key advantage of this architecture versus other approaches is that it reliably buffers power fluctuations without requiring any changes to system software, making it compatible with any training workload.
\end{abstract}

\begin{IEEEkeywords}
Datacenter power, active filtering, power electronics, energy storage, AI training, jitter, ramp rate control
\end{IEEEkeywords}

\input{1-intro}
\input{2-problem_formulation}
\input{3-overview_and_eval}
\input{4-converters}
\input{5-controls}
\input{6-discussion}
\input{7-conclusion}

\bibliographystyle{IEEEtran}
\bibliography{references}

\end{document}

%% file: 1-intro.tex
\section{Introduction}\label{sec:intro}
The advent of generative artificial intelligence (AI) has ushered in a new paradigm in both computing and the power grid. With large datacenter loads constituting an increasing share of electricity demand, the power grid faces unprecedented challenges in maintaining reliability~\cite{nerc2025largeloads}.

\begin{figure} 
    \centering
    \includegraphics[
        width=\columnwidth,
        trim=0 0.4cm 0 0, 
        clip
    ]{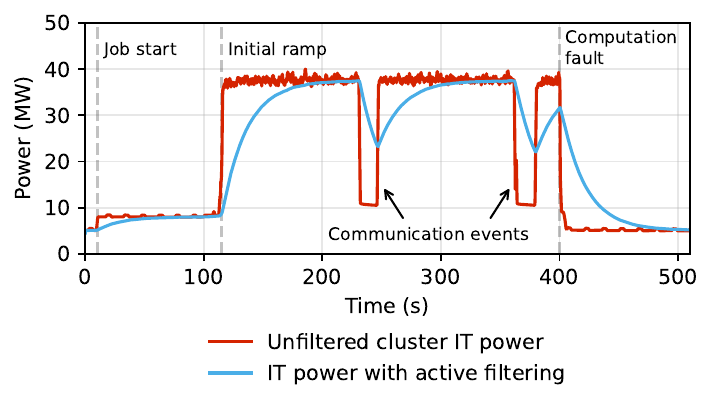}
    \caption{Unfiltered (red) and filtered (blue) power traces derived from a training job on NVIDIA H100 GPUs. Because the cluster works in unison during AI training, ramping events are amplified at scale. Initial ramp up, periodic communication events, and unpredictable faults all cause sudden ramping---in this instance, we measured rates as high as 193.7 MW/s (11.6 GW/min), which is far outside the range of what conventional generators could compensate for. The blue trace shows the effect of active filtering, which can confine the maximum ramp rate to within a prespecified limit ($\pm$2 MW/s in this example), without modifying underlying software.}
    \label{fig:power_at_scale}
\end{figure}

The scale of computation required to train a large AI model requires thousands of processors working synchronously, computing at high power with periodic low-power communication events~\cite{narayanan2021efficient, liunseen}. The fact that the entire cluster works in unison means that power ramps are amplified at scale. Figure~\ref{fig:power_at_scale} demonstrates this behavior in a power trace derived from a training job using NVIDIA H100 GPUs. Ramping between peak and idle power occurs at job startup, shutdown, communication/checkpointing, job failure, etc., and a single training job can last for weeks or months~\cite{li2025ai,choukse2025power}.

\input{figures/circuit-schematic}

The broader issue is that power grid stability depends on maintaining consistent balance between supply and demand for power. Grid operators dispatch and curtail capacity as needed to compensate for changing demand throughout the day, typically holding generators on standby to keep up with unexpected ramping events. Generators, however, are physical systems with inertia and thermal constraints that limit how fast they can ramp up or turn off. Depending on the generator, ramping times can take anywhere from a few seconds to a few hours. Bringing new generators online can take anywhere from a few minutes to a few days---a top-of-the-line 200 MW gas turbine used for load balancing may only be able to ramp at 0.5 MW/sec (0.03 GW/min)~\cite{abudu2021gas}. For comparison, the 40 MW cluster whose power trace is depicted in Figure~\ref{fig:power_at_scale} clocked peak-to-idle ramping rates as high as 193.7 MW/sec (11.6 GW/min), in the case of the fault-induced ramp-down event shown around t=400 seconds.

Analyses of recent transmission incidents have led to warnings of potential for cascading failures when available generators are not able to compensate for variable datacenter loads~\cite{nerc2024,lin2024exploding,ercot2025}. Step changes in power demand can excite torsional resonances that endanger spinning generators and other grid-connected systems~\cite{GE_torsional_dynamics,torsional_interaction, nerc2019oscillation}. In some areas, new datacenter projects may be denied unless the operators agree to stringent ramp rate restrictions~\cite{AESO2025datacentreconnection}.

This work proposes a hardware solution to the problem of AI-training power fluctuations in the form of a power supply equipped with an active filtering system. We propose to implement this solution on the rack level, rather than requiring campus-wide energy storage or balancing generators. A similar concept has been suggested previously in~\cite{choukse2025power}, but never defined rigorously or demonstrated in a physical system. We will show through prototype validation that this approach is a viable solution which guarantees that the rack power ramp rate will be confined to a prespecified range in compliance with essentially any grid-imposed restrictions.

The remainder of the paper is laid out as follows: Section~\ref{sec:problem_formulation} defines general operating restrictions that have to be met in order to protect the power grid from AI-induced power transients; Section~\ref{sec:overview} describes the different components of our active filter-equipped power system; Section~\ref{sec:hardware} describes the hardware and controller design for each subsystem in detail, along with the expected filter response of the assembled system;
Section~\ref{sec:discussion} discusses results from a built 10kW prototype system, exploring advantages and limitations of this specific design as well as areas where further improvements can be made; finally, Section~\ref{sec:conclusion} concludes with a summary of key findings.

%% file: figures/circuit-schematic.tex
\begin{figure*}[!b]
    \begin{center}
    \resizebox{0.95\linewidth}{!}{
    \centering
    \begin{circuitikz}[american voltages]
    \def\fcScale{0.8}
    \def\border{1*\fcScale}
    \def\dampingLine{4.5*\fcScale}
    \def\topLine{3.5*\fcScale}
    \def\bottomLine{0*\fcScale}
    \def\VDCx{-2*\fcScale}
    \def\filterLeftx{1*\fcScale}
    \def\filterRightx{5*\fcScale}
    \def\dampingCenterx{3*\fcScale}
    \def\CFx{5*\fcScale}
        \draw (\VDCx,\topLine) to [V=$V_{DC}$] (\VDCx,\bottomLine) 
            (\VDCx, \topLine) to [short, i_=$i_{DC}$](\filterLeftx-\border,\topLine) -- (\filterLeftx,\topLine)
            to [L=$L_F$] (\filterRightx,\topLine) 
            (\VDCx,\bottomLine)-- (\filterRightx,\bottomLine) to [C={$C_F$}] (\filterRightx,\topLine) 
            (\filterLeftx,\topLine) -- (\filterLeftx,\dampingLine)
            to [L=$L_{Da}$] (\dampingCenterx,\dampingLine) 
            to [R=$R_{Da}$] (\filterRightx,\dampingLine) 
            -- (\filterRightx,\topLine);
        \draw[dashed]
            (\filterLeftx-\border,\dampingLine+\border) rectangle (\filterRightx+\border,\bottomLine-\border);
        \node at (\dampingCenterx,\bottomLine-\border/2) {(a) Input filter};
    \def\VregLeftx{9*\fcScale}
    \def\VregRightx{10*\fcScale}
    \def\GPUsx{13*\fcScale}
    \def\GPUsYTop{2}
        \draw (\CFx,\topLine) -- (\VregLeftx,\topLine)
        ({(\CFx+\VregLeftx)/2},\topLine) to [open, v=$V_{IN}$]({(\CFx+\VregLeftx)/2},\bottomLine) 
        (\CFx,\topLine) to [short, i=$i_{IN}$](\VregLeftx,\topLine) 
        (\CFx,\bottomLine) -- (\VregLeftx,\bottomLine)
        (\VregRightx,\topLine) -- (\GPUsx,\topLine)
        (\VregRightx,\bottomLine) -- (\GPUsx,\bottomLine)
        (\GPUsx,\topLine) to [short,i=$i_{R}$](\GPUsx,\GPUsYTop) to [vR=$Rack$, invert](\GPUsx,\bottomLine) 
        ({(\VregRightx+\border+\GPUsx)/2},\topLine) to [open, v=$V_{OUT}$]({(\VregRightx+\border+\GPUsx)/2},\bottomLine);
        \draw[dashed]
            (\VregLeftx-\border,\topLine+\border) rectangle
            (\VregRightx+\border,\bottomLine-\border);
        \node[align=center] at 
            ({(\VregLeftx+\VregRightx)/2}, {(\topLine+\bottomLine)/2}) 
            {(b)\\DC-DC\\voltage\\regulator};
    \def\BidirectionalLeftx{16*\fcScale}
    \def\BidirectionalRightx{17*\fcScale}
    \def\Batx{19*\fcScale}
        \draw (\GPUsx,\topLine) to [short, i=$i_{B}$](\BidirectionalLeftx-\border,\topLine) -- (\BidirectionalLeftx,\topLine)
            (\GPUsx,\bottomLine) -- (\BidirectionalLeftx,\bottomLine)
            (\BidirectionalRightx,\topLine) -- (\Batx,\topLine)
            to [battery=$B_{AUX}$](\Batx,\bottomLine)
            -- (\BidirectionalRightx,\bottomLine);
        \draw[dashed]
            (\BidirectionalLeftx-\border,\topLine+\border) rectangle (\BidirectionalRightx+\border,\bottomLine-\border);
        \node[align=center] at
            ({(\BidirectionalLeftx+\BidirectionalRightx)/2}, {(\topLine+\bottomLine)/2}) {(c)\\Bidirectional\\converter};
    \end{circuitikz}
    }
\end{center}
    \caption{The system architecture consists of three main hardware components: (a) an input filter to remove high frequency power fluctuations and switching noise, (b) a DC-DC converter to provide a constant rack voltage, and (c) an auxiliary battery system to store or dispatch energy during transients. This configuration allows the power grid to gradually transition between different load conditions while the rack sees immediate power availability.}
    \label{fig:circuit-schematic}
\end{figure*}
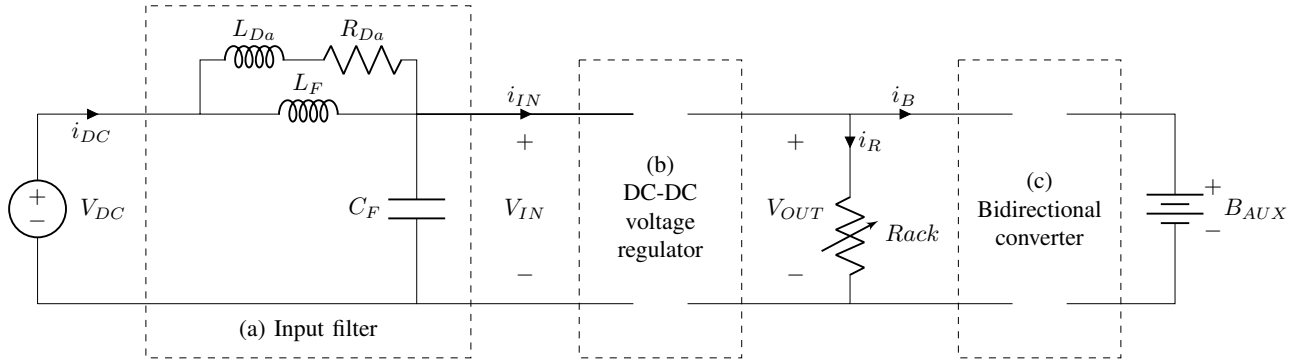

%% file: 2-problem_formulation.tex
\section{Framing the problem}\label{sec:problem_formulation}
In this work we consider a datacenter power trace ``well-behaved'' from the perspective of the broader power grid if it complies with the following rules:
\begin{enumerate}
    \item\label{item:ramping} The datacenter's electricity demand does not ramp faster than connected generator capacity can compensate.
    \item\label{item:frequency-content} The datacenter power trace does not induce resonance in power lines or connected hardware; it also must comply with broader regulations governing EMI and power quality.
\end{enumerate}

To ensure compliance with Rule~\ref{item:ramping} above, a utility might stipulate that a datacenter's load can't change by more than some specificied proportion of its total contracted capacity per second, as was the case in~\cite{AESO2025datacentreconnection}. Because the total load is the sum of all the individual electrical loads in the datacenter, we know for certain that a given AI training cluster will not break Rule~\ref{item:ramping} if every rack in the cluster individually obeys this rule, ramping up/down by at most the same specified proportion of its own rated power per second.

Low-pass filtering techniques are already broadly used to ensure EMI compliance, and the same principles can be extended to attenuate frequency content in forbidden resonance bands in order to comply with Rule~\ref{item:frequency-content}.

Different power grids will have different restrictions on ramp rate depending on the regional generation portfolio. Unfiltered AI training clusters will exhibit different power profiles depending on the nature of the job~\cite{mughees2025short}. In this work we propose a rack power supply architecture which can be easily adapted to ensure compliance with both ramp rate and frequency content requirements. The system we propose is in essence a low-pass filter comprised of both active and passive subsystems, keeping AI training racks in the ``well-behaved'' operating regime for any training load.







%% file: 3-overview_and_eval.tex
\section{System Overview}\label{sec:overview}

Figure~\ref{fig:circuit-schematic} depicts a simplified diagram for the proposed rack power supply unit.
Each subsystem carries out a specific function in meeting these goals:

\noindent\textbf{Input filter:}
The active components in the other subsystems introduce high-frequency switching harmonics that have to be attenuated for EMI compliance. The input filter heavily attenuates these harmonics, and if adequately sized it can also help attenuate lower frequencies corresponding to torsional resonances.

\noindent\textbf{DC-DC voltage regulator:}
Next-generation high-power racks would expect a constant $V_{OUT}=V_{DC}=400 V$, but $V_{IN}$ may rise or fall due to variations in $i_{IN}$~\cite{ocp_diablo400_v070}.
For example, with a large, poorly-damped second-order LC input filter, a step change ($\Delta~i_{IN}$) in load current will cause the capacitor voltage $V_{IN}$ to sinusoidally oscillate within the range
\begin{equation}
    V_{DC} - \Delta~i_{IN}\sqrt{\frac{L_F}{C_F}} \leq V_{IN} \leq V_{DC} + \Delta~i_{IN}\sqrt{\frac{L_F}{C_F}}
\end{equation}
assuming quiescent initial conditions. The DC-DC voltage regulator uses closed-loop compensation to maintain a constant voltage $V_{OUT}$ so that the rack maintains immediate power availability.

\noindent\textbf{Bidirectional converter:}
It is not reasonable to use a passive input filter alone to limit the rack's power ramp rate simply because of the sheer amount of energy that needs to be stored in order to buffer large power fluctuations over several seconds. The bidirectional converter interfaces with on-rack energy storage (in our system, a battery pack), storing or dispatching energy as needed to cancel out fluctuating rack power and limit the rack's maximum ramp rate.\footnote{This is similar to the role of the on-board traction battery in a hybrid car, where the system intentionally limits how fast the combustion engine can rev in order to maximize fuel econonomy. The battery in a hybrid compensates by discharging during acceleration and charging during regenerative braking, buffering the engine from sudden ramping events~\cite{ceraolo2016aging}.}

%% file: 4-converters.tex
\section{System Design \& Control}\label{sec:hardware}
We now describe the subsystem designs, their controllers, and the resulting overall filtering behavior.

\subsection{Cascaded voltage regulator design}\label{subsection:cascaded}
Figure~\ref{fig:DC-DC} shows the circuit topology we used in our prototype to regulate the voltage across the rack.
\input{figures/DC_Regulator}
A cascaded buck$\to$boost converter is ideal for this application because it has lower switch stress than a buck-boost- or Cúk-like topology. During normal operation, the floating switch $Q_{1a}$ is left on most of the time, and $Q_{2a}$ is generally turned off. This means the system incurs negligible switching losses when $V_{IN}=V_{OUT}$. The output voltage is regulated using pulse-width modulation---decreasing the positive duty cycle on $Q_{1a}$ causes a drop in $V_{OUT}$; increasing the duty cycle on $Q_{2a}$ raises $V_{OUT}$. Referring to these duty cycles as $D_{1a}$ and $D_{2a}$ respectively, the voltage gain ratio is approximately
\begin{equation}
\frac{V_{OUT}}{V_{IN}}\approx\frac{D_{1a}}{1-D_{2a}}
\end{equation}
when the converter is operated in the continuous conduction mode\footnote{i.e.\ the inductor always drives a positive current toward the load side of the converter.} (CCM).

\noindent\textbf{Regulator feedback controller design:}
For frequencies roughly an order of magnitude or more less than the switching frequency, the transient circuit characteristics can be modelled using a nonlinear average circuit approximation. This model can be used to analyze the small-signal model for a circuit and its corresponding transfer functions. For the sake of space this derivation is not performed here, but an interested reader can find a thorough description of the process in Chapters 7 and 8 of~\cite{erickson2020fpe}.
For the cascaded buck$\to$boost converter, we assume only one switch is modulated at a time. This leads to slightly different open-loop transfer functions depending on whether the converter is in buck or boost mode. When the converter starts with no net voltage gain (i.e. $V_{IN}=V_{OUT}$), the transfer functions are:\footnote{We use the convention that values with a tilde (e.g. $\tilde{v}_{OUT}$, $\tilde{d}$, etc.) indicate small-signal perturbations, while values without a tilde are either quiescent conditions or system constants.}
\begin{equation}
H_{OD1}(s)=\frac{\tilde{v}_{OUT}(s)}{\tilde{d}_{1a}(s)} \approx V_{IN}\frac{1}{1+\frac{s}{Q\omega_p}+\frac{s^2}{\omega_p^2}}
\label{equation:DC-buck-TF}
\end{equation}
\begin{equation}
H_{OD2}(s)=\frac{\tilde{v}_{OUT}(s)}{\tilde{d}_{2a}(s)} \approx V_{IN}\frac{1-\frac{s}{\omega_z}}{1+\frac{s}{Q\omega_p}+\frac{s^2}{\omega_p^2}}
\label{equation:DC-boost-TF}
\end{equation}
where $Q\approx\frac{V_{OUT}}{i_R+i_B}\sqrt{\frac{C_{DC}}{L_{DC}}}$, $\omega_p\approx\frac{1}{\sqrt{L_{DC}C_{DC}}}$, and $\omega_z\approx\frac{V_{OUT}}{(i_R+i_B)L_{DC}}$.

The stability criteria for feedback control stipulate that our controller transfer function $C(s)$ needs to interact with the open-loop system such that both $H_{OD1}(j\omega)\cdot C(j\omega)$ and $H_{OD2}(j\omega)\cdot C(j\omega)$ have a single crossover frequency and a ideally a phase margin of about 60°~\cite{beard2019loopshaping,erickson2020feedbacktheorem}.\footnote{Recall that the crossover frequency $\omega_{CO}$ of some transfer function $T(j\omega)$ refers to the frequency for which $|T(j\omega)| = 1 = 0 dB$, and the phase margin is how much the phase at the crossover frequency $\angle T(j\omega_{CO})$ differs from -180°.} The highest-frequency reference signal that a well-regulated closed-loop system can follow (i.e. the control bandwidth) is approximately the same as the crossover frequency, so ideally we want $\omega_{CO}$ to be as large as possible.

In our design, we implemented the controller using a fully analog feedback network. Figure~\ref{fig:DC-feedback} lays out the compensator circuit, where a PID compensator is implemented using passive components and an opamp with negative feedback. The inputs receive equally scaled readings of $V_{OUT}$ and $V_{DC}$. The output voltage $v_C$ serves as a reference signal that is shifted and compared against a ramp to generate pulsed gating signals for either $Q_{1a}$ or $Q_{2a}$ with essentially infinite duty cycle fidelity such that the steady-state error between $V_{OUT}$ and $V_{DC}$ falls to zero.
\input{figures/DC-DC-feedback}
In our prototype we used $C_{DC}=47 \mu F$, $L_{DC}=22 \mu H$, and selected feedback components for stable closed-loop control in both buck and boost modes: $R_1=16 \Omega$, $C_1=100 nF$, $R_2 = 630 \Omega$, $R_3 = 1.4 k\Omega$, and $C_3 = 110 nF$. We designed our converter to operate fully in CCM, delivering anywhere between 1 kW and 5 kW of regulated power, placing two identical converters in parallel for a total of up to 10 kW of regulated rack power.

\begin{figure}
    \centering
    \includegraphics[width=\linewidth]{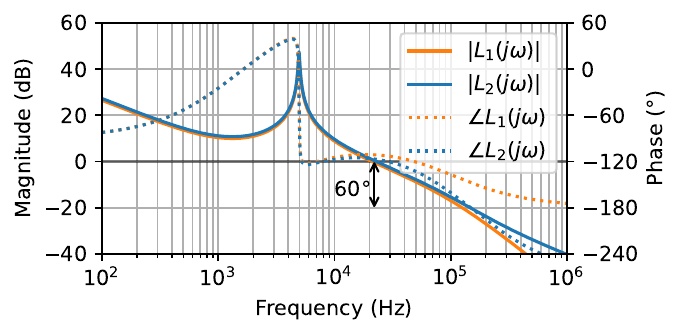}
    \caption{Closed-loop gain phases and magnitudes for the cascaded buck$\to$boost converter in Figure~\ref{fig:DC-DC} operating at 400 $V_{DC}$ and 5 kW load power, under PID regulation using the circuit shown in Figure~\ref{fig:DC-feedback}. Whether in buck or boost mode, the converter exhibits stable control dynamics with a single crossover frequency and roughly 60° phase margin.}
    \label{fig:DC-transfer-function}
\end{figure}

Figure~\ref{fig:DC-transfer-function} confirms that the loop gains $L_n(s)=H_{ODn}\cdot C(s)$ indicate stable control responses and a control bandwidth of approximately 20 kHz. (This is shown for a 5 kW load specifically, but the plot would look nearly identical for any load in the full operating range.)

\subsection{Bidirectional converter and active filter design}
Figure~\ref{fig:bidirectional} shows the bidirectional converter used to transfer energy between the rack busbar and the auxiliary battery pack. This buck- and boost-derived topology is one of the simplest bidirectional converter configurations, although alternative topologies may be preferred when galvanic isolation is required~\cite{Naidu2019DynamicEnergyManagement,karshenas2011bidirectional, alatai2021review,gorji2019topologies}.

\input{figures/bidirectional}

The prototype was designed to operate exclusively in the discontinuous conduction mode (DCM), such that the current through $L_B$ falls to zero before the end of each switching cycle. We need this system to be able to react very quickly to changing demand for auxiliary power. The DCM is appealing for this reason because of its fast dynamics, where a well-designed control system can enforce a given charge or discharge rate within a single switching cycle. The system can be held idle losslessly (whereas CCM battery management has switching, gating, and conduction losses at any power setpoint). DCM operation also allows us to use a wide range of duty cycles, leading to high fidelity when controlling charge/discharge rates, while in CCM the charge/discharge rate is ultimately set by the battery terminal resistance and only a narrow range of duty cycles are used. In DCM operation, the inductor $L_B$ acts as a built-in snubber, enabling zero-current switching at MOSFET turn-on and reducing switching losses. An additional snubber helps maintain high switch performance even though the turn-off transition is hard-switched.

A typical drawback to DCM operation is high current ripple observed on both the high- and low-voltage terminals of the converter. To minimize this effect, we used four parallel channels of interleaved bidirectional converters, each offset from the last by a quarter of a switching period. Because our system was designed for $V_{OUT}=V_{DC}=400 V$ and we used $V_B\approx100 V$, at maximum charge or discharge current, the four channels' ripples cancel out such that the battery sees negligible current ripple~\cite{wang2019switching,karshenas2011bidirectional,alatai2021review}.

Charging and discharging of the battery are controlled by pulse-width modulation (PWM) of either $Q_{1b}$ or $Q_{2b}$, respectively. Let $D^\star$ denote the duty cycle of the active switch, taken as the positive duty cycle ($D_{1b}$) when $Q_{1b}$ is switched and negative ($1-D_{2b}$) when $Q_{2b}$ is switched. The magnitude of the average current, $|i_B|$, transferred between the rack busbar and the battery through the sum of all $n$ converter channels is
\begin{equation}
|i_B|=n\cdot\frac{{\left(D^\star \right)}^2\left(V_{DC}-V_B\right)}{2L_B f_S}.
\label{eq:DCM-current}
\end{equation}
where $V_B$ is the battery terminal voltage, $f_S$ is the switching frequency, and we assume all $n$ channels are driven with the same duty cycle.

\noindent\textbf{Controller design for active filtering:}
The nature of the DCM is such that the transfer functions relating perturbations in $D_{1b}$ or $D_{2b}$ to changes in $i_B$ work out to be constant across all frequencies:\footnote{In equations~\ref{eq:DCM_charging_tf} and~\ref{eq:DCM_discharging_tf}, $\tilde{i}_B$ refers to the current out of or into the rack busbar respectively.}
\begin{equation}
    H_{ID1}(s)=\frac{\tilde{i}_B}{\tilde{d}_{1b}}=n\cdot\frac{D_{1b}}{f_S L_B}\left(V_{DC}-V_B\right)
    \label{eq:DCM_charging_tf}
\end{equation}
\begin{equation}
    H_{ID2}(s)=\frac{\tilde{i}_B}{\tilde{d}_{2b}}=n\cdot\frac{D_{2b}}{f_S L_B}\left(\frac{V_B^2}{V_{DC}-V_B}\right)
    \label{eq:DCM_discharging_tf}
\end{equation}
Consequently, a simple integral compensator is sufficient to ensure stability under closed-loop control. Figure~\ref{fig:bidirectional-feedback} lays out the circuit we used to implement both PI compensation and active filtering of rack power. 
\input{figures/bidirectional-feedback.tex}
The compensator output voltage $v_D$ is compared against a ramp to generate the variable duty cycle gating pulses for either $Q_{1b}$ or $Q_{2b}$. The net effect of the control system is that current is diverted to or from the battery system in opposition to changes in $i_R$.

In closed-loop operation, the compensator tracks the reference voltage $v_{REF}$, sending current to/from the battery at AC frequencies up to an upper limit loosely approximated by
\begin{equation}
\omega_H\approx\frac{1}{R_4C_4}
\label{eq:upper_cutoff}
\end{equation}

Because $v_{REF}$ in turn results from low-pass filtering of the rack current signal, the cutoff frequency
\begin{equation}
\omega_L=\frac{1}{R_5C_5}
\label{eq:lower_cutoff}
\end{equation}
sets the lower bound on the range of frequencies absorbed by the battery. 


If the battery system were to operate with perfect efficiency, it would only handle AC power and the battery's average state of charge (SoC) would be fixed. However, setpoint bias in the control system, converter inefficiencies, and round-trip inefficiency due to battery dynamics (i.e. current-dependent overvoltages, loss of active material, ohmic losses, etc.) all contribute to SoC drift over hours to weeks of system uptime~\cite{Rand_BatteriesForEVs,guena2006depth}. In our prototype, we implemented an auxiliary software system which interfaces with the battery management system to track SoC during system runtime. The software uses a predictive control scheme to dynamically issue corrective trickle charging commands, holding the SoC at a given setpoint with enough headroom to keep the battery from over/undervoltage and minimize calendar aging. This operation is described in greater detail in~\cite{jensen2026easyrider}.

\subsection{Input filter design}
As mentioned in Section~\ref{sec:overview}, the input filter's primary purpose is to attenuate switching harmonics for EMI compliance. It also helps to attenuate power fluctuations with frequencies too fast for the battery system to compensate.

In our prototype we used the second-order input filter topology shown in Figure~\ref{fig:circuit-schematic}(a). Fluctuations in $i_{IN}$  are attenuated by as much as 40 dB per order of magnitude above the corner frequency $\omega_F=1/\sqrt{L_F C_F}$. $L_F$ and $C_F$ are selected such that the system stays within the utility-defined harmonic content limits under worst-case power fluctuation conditions. The damping leg needs to be carefully designed in order to preserve the stability of the other subsystems' transfer functions. A thorough explanation of the relevant stability conditions can be found in~\cite{riccobono2014comprehensive} and chapter 17 of~\cite{erickson2020fpe}.

\subsection{Full system active filter effect and ramp rate limit}
\begin{figure}
    \centering
    \includegraphics[width=\columnwidth]{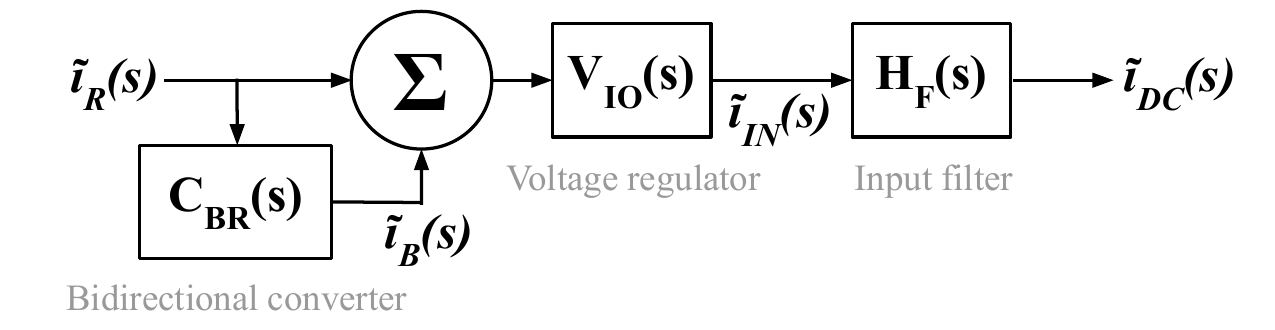}
    \caption{System current flow diagram, with transfer functions corresponding to the closed-loop behavior of each of the regulators and the input filter. The current sourced from the grid to power the fluctuating rack current results from the combined effect of the bidirectional converter, voltage regulator, and input filter subsystems.}
    \label{fig:signal-flow-diagram}
\end{figure}

Figure~\ref{fig:signal-flow-diagram} shows how each of the subsystems contributes to the net filtering effect of the system. We use $C_{BR}(s)=\frac{\tilde{i}_B(s)}{\tilde{i}_R(s)}$, $V_{IO}(s)=\frac{\tilde{i}_{IN}(s)}{\tilde{i}_R(s)+\tilde{i}_B(s)}$, and $H_F(s)=\frac{\tilde{i}_{DC}(s)}{\tilde{i}_{IN}(s)}$ to describe the individual transfer functions of each subsystem.\footnote{Because in our system $V_{OUT}=V_{DC}$, the transfer functions are the same when analyzing the system frequency response in terms of current or power.} Rack power fluctuations are attenuated by the combined effect of the whole power supply system before being observed on the grid side:
\begin{equation}
\frac{\tilde{i}_{DC}(s)}{\tilde{i}_R(s)}=\left(1+C_{BR}(s)\right)\left(V_{IO}(s)\right)\left(H_F(s)\right)
\label{eq:full_TF_exact}
\end{equation}

The complete control dynamics of the bidirectional converter cause frequency content between $\omega_L$ and $\omega_H$ to be absorbed in the battery. In context of the full system filter effect, this system's band-stop filter response can be approximately defined by
\begin{equation}
1+C_{BR}(s)\approx 1 - \frac{1}{\left(1+\frac{\omega_L}{s}\right)\left(1+\frac{s}{\omega_H}\right)}
\label{eq:1_plus_CBR}
\end{equation}

Because the cascaded regulator holds the output voltage constant for all frequencies lower than a few thousand Hz,
\begin{equation}
V_{IO}(s)\approx 1
\label{eq:VIO}
\end{equation}

Finally, the input filter's transfer function follows from current division between the inductive/damping branchs and capacitive branch depicted in Figure~\ref{fig:circuit-schematic}(a):
\begin{equation}
    H_F(s)=\frac{\frac{1}{sC_F}}{\frac{1}{sC_F}+(sL_D+R_D)||sL_F}
\label{eq:HF}
\end{equation}

\begin{figure}
    \centering
    \includegraphics[width=\columnwidth]{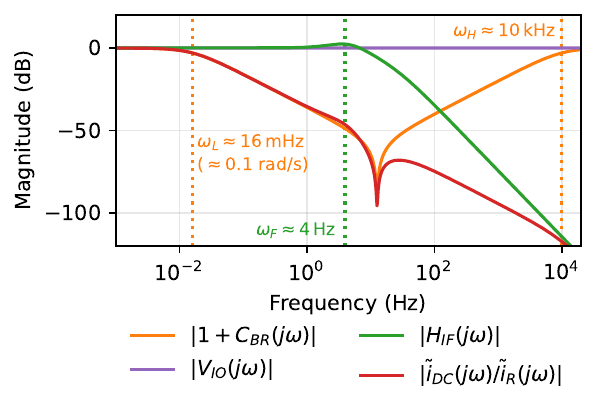}
    \caption{Subsystem transfer functions (orange, green, purple) and the full system's frequency response (red). The battery system selectively filters power transients between $\omega_L$ and $\omega_H$, while the passive filter contributes progressive attenuation for all frequencies above $\omega_F$.}
    \label{fig:full-transfer-function}
\end{figure}

Figure~\ref{fig:full-transfer-function} plots the magnitudes of all three subsystem transfer functions along with the full-system frequency response. As expected, fluctuations in rack power at all frequencies above $\omega_L$ are attenuated before propagating out to the DC bus.
For the system described in Figure~\ref{fig:full-transfer-function}, we used $R_4=1.6k\Omega$, $C_4=10nF$, $R_5=100k\Omega$, $C_5=100\mu F$, $L_F=54mH$, $C_F=30mF$, $L_D=4.7mH$, and $R_D=1\Omega$, with the voltage regulator system identical to the description in sub-section~\ref{subsection:cascaded}.

Although the battery system's closed-loop filtering effect tapers off as frequencies approach $\omega_H$, the passive input filter's corner frequency is deliberately placed before the inflection point in $1+C_{BR}(s)$. As a result, the full system transfer function is at least as effective as a first-order low-pass filter with a cutoff frequency of $\omega_L$:
\begin{equation}
    \left|\frac{\tilde{i}_{DC}(s)}{\tilde{i}_R(s)}\right| \leq \left|\frac{1}{1+\frac{s}{\omega_L}}\right|
    \label{eq:full_TF_approx}
\end{equation}

Fourier analysis of AI-training job power traces reveals that nearly all high-amplitude power fluctuation frequency content is concentrated in frequencies below about 1 Hz~\cite{choukse2025power}. Because the passive input filter has no effect for frequencies this low, the full system ramp rate is really only limited by the dynamics of the battery system. In this frequency range the compensating current $i_B$ can be approximated as
\begin{equation}
i_B(s)\approx-\frac{1}{1+\frac{\omega_L}{s}}i_R(s)
\label{eq:ib_freq}
\end{equation}
or using the equivalent time-domain differential equation
\begin{equation}
\frac{d}{dt}i_B(t)+\omega_L i_B(t)+\frac{d}{dt}i_R(t)\approx0
\label{eq:ib_time}
\end{equation}

Assuming $i_B(t)$ stays within predictable limits, let's define
\begin{equation}
\alpha=\frac{P_{RATED}-P_{IDLE}}{P_{RATED}}
\label{eq:alpha}
\end{equation}
where $P_{RATED}$ is the maximum rated rack power, and $P_{IDLE}$ is the minimum rack power.
It follows from (\ref{eq:ib_time}) that the combined ``load'' power $P_L(t)=V_{DC}\left(i_R(t)+i_B(t)\right)\approx V_{DC}\cdot i_{DC}(t)$  has an absolute limit on its ramp rate, defined by
\begin{equation}
\left|\frac{d}{dt}P_L(t)\right|\leq\omega_L\cdot\alpha P_{RATED}
\label{eq:ramp_limit}
\end{equation}

In other words, the system's maximum ramp rate is set by the cutoff frequency $\omega_L$, even without any modifications to the software actually running on the rack.

\subsection{Battery sizing}
It is essential to know the maximum energy that might need to be stored in or sourced from the battery in order to implement this system in practice. Fortunately, this problem has a closed-form solution.

Suppose the battery is initially at rest at time $t=0$ with the rack drawing power $P_1$ from the grid. Over some period of time, the rack transitions to drawing power $P_2$ and stays there. The net change in battery energy in this scenario is
\begin{equation}
\Delta E_B=V_{DC}\int_{0}^{\infty}i_B(t)dt
\label{eq:energy_ib}
\end{equation}
Rearranging (\ref{eq:ib_time}) to solve for $i_B$, (\ref{eq:energy_ib}) becomes
\begin{equation}
\Delta E_B=-\frac{V_{DC}}{\omega_L}\int_{0}^{\infty}\left(\frac{d}{dt}\left(i_R(t)+i_B(t)\right)\right)dt
\label{eq:energy_ir_ib}
\end{equation}
From (\ref{eq:ib_time}) we can deduce that $i_B$ decays exponentially to zero during periods where $i_R$ is constant. Therefore,
\begin{equation}
\Delta E_B = -\frac{V_{DC}}{\omega_L}\left[i_R(t)+i_B(t)\right]_{t=0}^{t=\infty}=\frac{P_1-P_2}{\omega_L}
\label{eq:energy_int}
\end{equation}
From (\ref{eq:ib_time}) we know that the battery won't ever charge or discharge unless the rack power has generally decreased or increased, respectively. The maximum net energy displacement therefore occurs when the rack transitions between $P_{RATED}$ and $P_{IDLE}$ and holds long enough at each value for the battery current to settle to zero. It follows that for any possible rack power trace, the net energy stored or discharged from the battery is bounded at all times $t\in[0,\infty)$ by
\begin{equation}
    |\Delta E_B| = |V_{DC}\int_{0}^{t}i_B(\tau)d\tau| \leq \frac{\alpha}{\omega_L}P_{RATED}
    \label{eq:energy_absolute}
\end{equation}
where again $\alpha$ is as defined in (\ref{eq:alpha}), and $\omega_L$ is defined in (\ref{eq:lower_cutoff}).

Consequently, the right hand side of (\ref{eq:energy_absolute}) expresses the minimum possible capacity that an auxiliary energy storage system would need in order to buffer an arbitrary waveform using the control dynamics described by (\ref{eq:ib_time}). However, using a battery only this large would require 100\% depth of discharge (DoD) to buffer long power transients---this is not ideal given that higher DoD per cycle correlates with significant battery aging acceleration~\cite{soto2022impact,takei2001cycle,gantenbein2019capacity,guena2006depth}. For this reason we add a final term, $\gamma$, denoting the proportion of battery capacity that is allowed to be used in buffering arbitrary transients. The minimum battery capacity that meets this requirement is
\begin{equation}
    E_{MIN}\geq \frac{\alpha}{\gamma\omega_L}P_{RATED}
    \label{eq:energy_minimum}
\end{equation}

Finally, the battery must be rated to absorb the full difference between maximum and minimum rack power. The minimum rated battery power, $P_B$ is
\begin{equation}
P_B\geq\alpha P_{RATED}
\label{eq:power_minimum}
\end{equation}


%% file: figures/DC_Regulator.tex
\begin{figure}[b]
    \centering
    \begin{circuitikz}[american voltages, scale=0.92, transform shape]
        \def\fcScale{0.75}
        \def\border{1*\fcScale}
        \def\diodeScale{0.6}

        \def\topLine{3.5*\fcScale}
        \def\bottomLine{0*\fcScale}
        \def\outerx{-1*\fcScale}
        \def\leftx{0*\fcScale}
        \def\buckrightx{3*\fcScale}
        \def\boostleftx{6*\fcScale}
        \def\boostrightx{9*\fcScale}
        \def\voutx{10.5*\fcScale}
        \def\rightx{11.5*\fcScale}

        \draw(\outerx, \bottomLine) -- (\rightx,\bottomLine);

        \draw(\outerx,\topLine) to [short, i=$i_{IN}$] (\leftx,\topLine)
            to [open, v=$V_{IN}$] (\leftx,\bottomLine);

        \draw($( \leftx,\topLine )!0.5!( \buckrightx,\topLine )$)
            node[nigfete, bodydiode, rotate=90] (Q1) {}
            (Q1.D) -- (\leftx,\topLine)
            (Q1.S) -- (\buckrightx,\topLine);
        \node[left=-3pt] at (Q1.G) {$Q_{1a}$};

        \begin{scope}[transform shape, scale=\diodeScale]
            \draw (\buckrightx / \diodeScale,\bottomLine / \diodeScale)
                to [D] (\buckrightx / \diodeScale,\topLine / \diodeScale);
        \end{scope}
        \node[right=4pt] at ($( \buckrightx,\bottomLine)!0.48!( \buckrightx,\topLine )$) {$D_1$};

        \draw (\buckrightx,\topLine) to [L, l_=$L_{DC}$] (\boostleftx,\topLine);

        \draw($(\boostleftx,\topLine)!0.5!(\boostleftx,\bottomLine)$)
            node[nigfete, bodydiode] (Q2) {}
            (Q2.D) -- (\boostleftx,\topLine)
            (Q2.S) -- (\boostleftx,\bottomLine);
        \node[left=1pt, yshift=3pt] at (Q2.S) {$Q_{2a}$};

        \begin{scope}[transform shape, scale=\diodeScale]
            \draw (\boostleftx / \diodeScale,\topLine / \diodeScale)
                to [D] (\boostrightx / \diodeScale,\topLine / \diodeScale);
        \end{scope}
        \node[below, yshift=-3pt] at ($( \boostrightx,\topLine)!0.48!( \boostleftx,\topLine )$) {$D_2$};

        \draw(\boostrightx,\topLine) to [curved capacitor, l_=$C_{DC}$]
            (\boostrightx,\bottomLine);

        \draw(\boostrightx,\topLine) -- (\rightx,\topLine);
        \draw(\voutx,\topLine) to [open, v=$V_{OUT}$] (\voutx,\bottomLine);
        \draw(\voutx,\topLine) to [short, i=$i_R + i_B$] (\rightx,\topLine);

    \end{circuitikz}

    \caption{Cascaded buck$\to$boost DC-DC converter topology for the rack voltage regulator referenced in Fig.~\ref{fig:circuit-schematic}(b). Gate drive circuitry, snubbers, etc. ommitted for clarity.}
    \label{fig:DC-DC}
\end{figure}

%% file: figures/DC-DC-feedback.tex
\begin{figure}
    \begin{center}
        \def\Scale{0.8}
        \begin{circuitikz}[american, scale=\Scale, transform shape]

            \def\topY{2}
            \def\midX{3.5}
            \def\rightX{7.0}

            \node[op amp] (opamp) at (6,0) {};

            \draw (opamp.+) -- ++(-0.8,0)
                node[circ]{}
                node[left] {$k\cdot V_{DC}$};

            \coordinate (midref) at (0,0 |- opamp.-);

            \coordinate (sum) at (\midX,0 |- opamp.-);
            \draw (sum) node[circ] {};
            \draw (sum) -- (opamp.-);

            \coordinate (vs) at (0,0 |- opamp.-);
            \draw (vs) -- (-0.5,0 |- opamp.-) node[circ]{} node[left]{$k\cdot V_{OUT}$};

            \draw
                (vs) -- (0,\topY)
                to[R=$R_2$] (\midX,\topY)
                -- (sum);

            \draw
                (vs)
                to[R=$R_1$] (0.5*\midX,0 |- opamp.-)
                to[C=$C_1$] (sum);

            \draw
                (\midX,\topY)
                to[R=$R_3$] ($(\midX,\topY)!0.5!(\rightX,\topY)$)
                to[C=$C_3$] (\rightX,\topY)
                -- (\rightX, 0 |-opamp.out);

            \draw
                (opamp.out) -- ++(0.5,0)
                node[circ]{}
                node[right] {$v_C$};

        \end{circuitikz}
    \end{center}

    \caption{Op amp compensator circuit, implementing PID control such that $V_{OUT}$ tracks $V_{DC}$ despite changing load currents. The value $k$ is a constant indicating the attenuation factor when stepping the output voltage down to a range compatible with the compensator circuit. The voltage $v_C$ serves as a reference voltage that sets the duty cycles $D_{1a}$ and $D_{2a}$}
    \label{fig:DC-feedback}
\end{figure}

%% file: figures/bidirectional.tex
\begin{figure}
    \begin{center}
    \centering
        \begin{circuitikz}[american voltages]
            \def\fcScale{0.8}

            \def\topLine{5*\fcScale}
            \def\middleLine{2.5*\fcScale}
            \def\bottomLine{0*\fcScale}
            \def\outerx{-2.2*\fcScale}
            \def\leftx{0*\fcScale}
            \def\rightx{3*\fcScale}

            \draw (\outerx,\bottomLine) -- (\rightx,\bottomLine);

            \draw (\outerx,\topLine) to [short, i=$i_B$](\leftx,\topLine);
            \draw(\outerx,\topLine) to [open, v=$V_{OUT}$](\outerx,\bottomLine);

            \draw ($( \leftx,\topLine )!0.5!( \leftx,\middleLine )$)
                node[nigfete, bodydiode] (Q1) {};
            \draw (Q1.D) -- (\leftx,\topLine);
            \draw (Q1.S) -- (\leftx,\middleLine);
            \node[left=1pt, yshift=3pt] at (Q1.S) {$Q_{1b}$};

            \draw ($( \leftx,\middleLine )!0.5!( \leftx,\bottomLine )$)
                node[nigfete, bodydiode] (Q2) {};
            \draw (Q2.D) -- (\leftx,\middleLine);
            \draw (Q2.S) -- (\leftx,\bottomLine);
            \node[left=1pt, yshift=3pt] at (Q2.S) {$Q_{2b}$};

            \draw (\leftx,\middleLine) to [L, l_=$L_B$] (\rightx,\middleLine);

            \draw (\rightx,\middleLine) to [battery, l=$V_{B}$] (\rightx,\bottomLine);

        \end{circuitikz}
    \end{center}
\caption{A single channel of the bidirectional converter as referenced in Fig.~\ref{fig:circuit-schematic}(c), used to charge/discharge the battery as needed to mitigate fluctuations in $i_{R}$. Gate drive circuitry, snubbers, etc. ommitted for clarity.}
\label{fig:bidirectional}
\end{figure}

%% file: figures/bidirectional-feedback.tex
\begin{figure}
    \begin{center}
        \def\Scale{0.8}
        \begin{circuitikz}[american, scale=\Scale, transform shape]

            \def\topY{1.5}
            \def\midX{2.5}
            \def\rightX{5.1}

            \node[op amp] (opamp) at (4,0) {};

            \coordinate (midref) at (0,0 |- opamp.-);

            \coordinate (sum) at (\midX,0 |- opamp.-);
            \draw (sum) node[circ] {};
            \draw (sum) -- (opamp.-);

            \coordinate (vs) at (0,0 |- opamp.-);
            \draw (vs) -- (0,0 |- opamp.-) node[circ]{} node[left]{$g\cdot i_B$};

            \coordinate (ir) at (0,0 |-opamp.+);
            \draw (ir)
                node[circ]{}
                node[left] {$-g\cdot i_R$};
            \draw (ir)
                to[C, l_=$C_5$] (\midX, 0|-opamp.+) -- ++(0,-0.5)
                to[R=$R_5$] ++(0,-1.5)
                node[ground]{};
            \draw (\midX, 0|-opamp.+) node[circ, label=above:$v_{\mathrm{REF}}$] {} -- (opamp.+);

            \draw (vs) to[R=$R_4$] (sum);

            \draw
                (sum) -- (\midX,\topY)
                to[C=$C_4$] (\rightX,\topY)
                -- (\rightX, 0 |-opamp.out);

            \draw
                (opamp.out) -- ++(0.5,0)
                node[circ]{}
                node[right] {$v_D$};

        \end{circuitikz}
    \end{center}

    \caption{Op amp compensator circuit, implementing PI control such that the closed-loop current $i_{B}$ opposes changes in rack current $i_R$. The value $g$ is a constant indicating the scale factor from the circuits used to measure current. The voltage $v_D$ serves as a reference voltage that sets the duty cycles $D_{1b}$ and $D_{2b}$. Component values $C_5$ and $R_5$ determine the lower limit of frequencies absorbed by the battery system and therefore the full system ramp rate.}
    \label{fig:bidirectional-feedback}
\end{figure}

%% file: 5-controls.tex

%% file: 6-discussion.tex
\section{Results \& Discussion}\label{sec:discussion}
We constructed a prototype system rated for up to 10 kW of buffered power at 400 $V_{DC}$ according to the design methodology and part values laid out in Section~\ref{sec:hardware}, where the maximum ramp rate is designed to be $\omega_L=0.1$ p.u./s. We built a custom 74 Ah 100 V battery pack using high-power LiFePO$_4$ cells. The system bill of materials cost came out to about \$3,500 USD, including some expenses associated with iteration and testing.\footnote{A full system photo was previously published in~\cite{jensen2026easyrider} and is not inlcuded here for the sake of space.}

\begin{figure}
    \centering
    \includegraphics[width=\columnwidth]{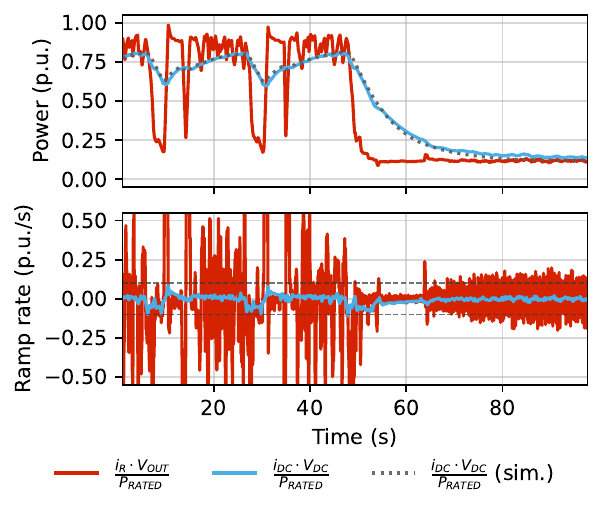}
    \caption{Top: Measured (blue) and simulated (dotted grey) power drawn from the system DC bus using a 10 kW built prototype and a simplified dynamics model, respectively. The unfiltered load power trace is shown in red. Bottom: measured ramp rates over the same timespan for as observed at the rack (red) and DC input bus (blue) terminals in the built prototype. The measured prototype never exceeds the expected maximum ramp rate of ±0.1 p.u./s.}
    \label{fig:full-demonstration}
\end{figure}

Figure~\ref{fig:full-demonstration} shows measured values from a test using the built prototype to power a programmable DC load in a laboratory setting compared with simulated values produced by approximating the composite system filtering effect as a first-order low-pass filter with a cutoff frequency of $\omega_L=0.1$ rad/s ($\approx16$ mHz).
The measured values shown in blue match the simulated values with remarkable precision. We observe the measured ramp rate never exceeds 0.1 p.u./s, despite high ramp rates in the unfiltered trace. This serves to validate the claim that we can expect a bounded ramp rate according to the relationship defined in (\ref{eq:ramp_limit}). 

\begin{figure}
    \centering
    \includegraphics[width=\columnwidth]{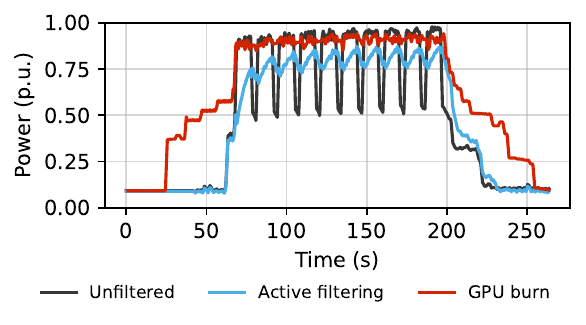}
    \caption{A comparison of the measured power traces resulting from unfiltered training (grey), training using active filtering in hardware as proposed in this work (blue), and software-based load-leveling using GPUS burns to fill any predictable dips in IT power (red). The traces in this figure are derived from training a GPT-like model on NVIDIA TITANX server.}
    \label{fig:titanx_comp}
\end{figure}

Using an active filter in this application makes it easy to adapt the system for compliance with grid-imposed ramp rate limits: rather than needing complex modifications to underlying software, the maximum system ramp rate is fixed by a single parameter ($\omega_L$) in the control loop. Figure~\ref{fig:titanx_comp} demonstrates how software-based load smoothing compares to active filtering in another demonstration of the prototype. In the ``GPU burn'' case shown, the ramp rate is limited by preprofiling the job at runtime and inserting random matrix multiplications during communication events and at startup and shutdown. In this example, the software-based approach used 19\% more energy than in-hardware active filtering. A hardware solution is also more reliable than a software one given that computation faults are generally not predictable\cite{liunseen,mughees2025short}.

Next-generation racks will already have significant energy storage onboard for uninterruptible power supply (UPS) in the case of a grid-side outage~\cite{ocp_diablo400_v070}. In a future rendition of this design, it may be better to use a digital control system. This would allow for consolidated integration of UPS functionality with the proposed active power filter hardware.

Battery degradation under constant cycling is an important consideration for this design. There is evidence to suggest that an oversized battery would perform significantly longer under constant shallow charging and discharging than a minimum-viable-capacity battery~\cite{soto2022impact,guena2006depth}. Further experimentation is needed to characterize and optimize battery lifetime. A hybrid energy storage system leveraging the complementary characteristics of supercapacitors or flywheels in addition to chemical batteries may improve system lifetime.

With some minor adjustments, the order of the subsystems could be swapped around such that the battery system only operates on power fluctuations that have already been attenuated by the large input filter currently depicted as subsystem (a) in Figure~\ref{fig:circuit-schematic}. This would alleviate some battery stress without affecting the form of the system filter response in (\ref{eq:full_TF_approx}).

%% file: 7-conclusion.tex
\section{Conclusion}\label{sec:conclusion}
In this work we laid out a design methodology for a power supply which uses active filtering to predictably limit the ramp rate of a single rack.
If each individual rack in a cluster were equipped with a power supply employing active filtering as described in this work, the total load would also behave with predictable limits on its absolute ramp rate. A scaled system ramp rate would be constrained by the relationship in (\ref{eq:ramp_limit}) in the same way a single rack or GPU would be, taking $P_{RATED}$ as the corresponding maximum system power.

We have described the function of each subsystem and provided key design equations. Equations (\ref{eq:energy_minimum}) and (\ref{eq:power_minimum}) describe in a closed form the minimum viable specifications for an energy storage system given a specified ramp rate limit. Equation (\ref{eq:full_TF_exact}) describes the complete system filtering response, which performs at least as well as the simplified expression (\ref{eq:full_TF_approx}). We validated this design on a prototype built to deliver 10 kW of buffered power at 400 $V_{DC}$.

The functionality of this design is entirely independent of underlying training software. By simply changing the way that we deliver power to datacenter loads, we can ensure that they will be ``well-behaved'' in the context of the broader power grid.